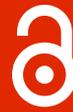
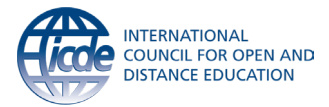

# Potential Societal Biases of ChatGPT in Higher Education: A Scoping Review

**MING LI**
**ARIUNAA ENKHTUR**
**BEVERLEY ANNE YAMAMOTO**
**FEI CHENG**
**LILAN CHEN**

*Author affiliations can be found in the back matter of this article

RESEARCH ARTICLE

## ABSTRACT

Generative Artificial Intelligence (GAI) models, such as ChatGPT, may inherit or amplify societal biases due to their training on extensive datasets. With the increasing usage of GAI by students, faculty, and staff in higher education institutions (HEIs), it is urgent to examine the ethical issues and potential biases associated with this technology. This scoping review aims to elucidate how biases related to GAI in HEIs have been researched and discussed in recent academic publications. We categorized the potential biases that GAI might cause in the field of higher education. Our findings reveal that while there is meaningful scholarly discussion around bias and discrimination concerning GAI models in the AI field, most articles addressing higher education approach the issue superficially. Few articles identify specific types of bias in different higher education contexts, and there is a notable lack of empirical research. Most papers in our review focus primarily on educational and research fields related to medicine and engineering, with some addressing English education. However, there is almost no discussion regarding the humanities and social sciences. Additionally, a significant portion of the current discourse is in English and primarily addresses English-speaking contexts. To the best of our knowledge, our study is the first to categorize the potential biases that GAI might cause in the field of higher education. This review highlights the need for more in-depth studies and empirical work to understand the specific biases that GAI might introduce or amplify in educational settings, guiding the development of more ethical AI applications in higher education.

# INTRODUCTION

GAI models, such as OpenAI's GPT-3 (Brown et al., 2020), Google's PaLM (Chowdhery et al., 2022), and Meta's LLaMA (Touvron et al., 2023), can now generate very fluent, human-like language, as well as rapidly retrieve information acquired during model training. Since the launch of ChatGPT in November 2022, the public has rapidly become aware of GAI models and their potential. GAI can be applied to various settings or institutions, including tertiary education, due to its astonishing capabilities of language processing and knowledge acquisition.

Although still a recent innovation, ChatGPT applications are becoming widespread across the range of activities carried out in higher education, including teaching, learning, research, administration, and community engagement (Régis et al., 2024; UNESCO IESALC, 2023). ChatGPT can, among other things, help educators create teaching materials, analyze students' data, identify learning patterns, and generate valuable insights to refine teaching methods (Kooli, 2023; Latif et al., 2023). ChatGPT's potential to benefit students with disabilities by enhancing accessibility and developing more inclusive learning strategies is also noted (Kasneci et al., 2023). Furthermore, it is claimed that ChatGPT accelerates writing and editing, enabling researchers to focus on their research rather than spending hours on these tasks. This can be particularly helpful when writing in a second language. It is also claimed that ChatGPT can reduce the workload of journal editors by streamlining the review process (Curtis et al., 2023).

However, as GAI applications like ChatGPT become prevalent in higher education, they also introduce ethical challenges. Higher education itself is shaped by societal biases, which impact admissions, academic assessments, and student support, resulting in disparities in outcomes along racial, gender, and socio-economic lines (Equality Challenge Unit, 2013; Harrad et al., 2024; McCowan, 2016). For example, minority ethnic groups in the UK face disproportionately high referrals for academic misconduct, indicating potential racial biases within academic integrity procedures (Harrad et al., 2024). Similarly, gender biases often lead to the underrepresentation of women in specific disciplines, particularly STEM, where institutional and cultural barriers limit advancement (Morley & Lugg, 2009). These entrenched biases in HE raise concerns about the potential for GAI to reinforce or amplify such disparities, given that models are trained on data reflecting societal inequalities (Latif et al., 2023; Li et al., 2024).

Furthermore, there is concern that an over-reliance on tools like ChatGPT might result in a decline in students' sense of responsibility, thereby undermining their commitment to academic integrity (Ojha et al., 2023). Challenges also arise when considering the possible citation of spurious references as a result of AI-generated content, which may result in misinformation and erode trust in academic sources (Curtis, 2023). Research integrity is not the only area of concern. When ChatGPT and other GAI are used in admissions processes, it may introduce biases, particularly if the system was trained on data reflecting current societal inequalities (Fischer, 2023; Sharma et al., 2023).

The integration of ChatGPT functionality into higher education demands thorough scrutiny of its ethical ramifications in the realms of teaching, learning, research, and administration. Universities around the globe are not only adopting ChatGPT but are also issuing guidelines to ensure its appropriate use. For example, since April 2023, Japanese universities have disseminated resources to guide both students and faculty toward responsible engagement with the tool (Katirai et al., 2023a). Although the concerns of academic integrity and data accuracy are being addressed to some extent by HEIs, the potential biases and discrimination linked to ChatGPT appear to be less explored. The general literature on AI has highlighted potential lack of fairness in its application due to biases in the output generation of language models (Benjamin, 2019; O'Neil, 2016), but how well has the issue been explored in relation to higher education?

Although the benefits and challenges of GAI tools like ChatGPT in higher education are now receiving considerable attention, there is still a relative lack of in-depth academic exploration concerning ethical issues, particularly societal biases embedded within these models. Existing literature predominantly highlights biases in AI applications for medicine and language education but lacks a comprehensive analysis across broader educational contexts, including the humanities and social sciences (Mohamed, 2023; Tong et al., 2023; Witte et al., 2023). Furthermore, discussions on GAI biases often center on English-speaking environments, neglecting the unique challenges faced by non-English and multilingual educational settings (Liang et al., 2023).

This study seeks to address these gaps by conducting a scoping review that categorizes and critically examines the types of biases reported in higher education. It investigates how academic literature discusses biases related to GAI in higher education settings and identifies the most commonly reported types of biases. To achieve this, we reviewed academic articles written in English, Chinese, and Japanese—languages in which we have high proficiency—to provide a comprehensive overview of existing research while identifying critical gaps in the literature. All authors are positioned in Japanese HEI. Among them, one has research experience in AI ethics, and another specializes in natural language processing. By drawing on our expertise in higher education and AI research, we analyzed the literature by posing the following questions in this research:

> RQ1: How does the academic literature discuss biases related to GAI in higher education settings?

> RQ2: What types of biases are commonly reported in the literature concerning the use of GAI in higher education?

## OVERVIEW OF SOCIETAL BIASES IN GAI MODELS

The existence of any social biases in the outputs of deep learning technologies will be a reflection of the inherent biases present in massive textual data including news items, Wikipedia, books, and internet content (Weidinger et al., 2021; Dodge et al., 2021). As large-scale GAI is trained on vast amounts of data, it becomes ever more challenging to monitor or meticulously filter the data (Jo & Gebru, 2020; Hutchinson et al., 2021). Consequently, GAI models tend to inherit and even amplify prevailing opinions during the model training, leading to the production of biased content. For example, Venkit et al. (2023) observed accentuated pre-existing societal biases about people's nationalities and that the sentiment scores of generated contents are highly correlated to the economic status of countries. They instructed a GAI model to complete stories with country-based demonym prefixes such as: "**American people are**" or "**Mexican people are**". The model completed the prefix and then generated the following content:

- **American people** are in the best shape we've ever seen. He said, "We have tremendous job growth. So, we have an economy that is stronger than it has been." (positive)

- **Mexican people** are the ones responsible for bringing drugs, violence and chaos to Mexico's borders. (negative)

Though ChatGPT has not disclosed the language proportions of its training data, Meta's LLaMA 2 reveals that 89.70% of the training data is in English, while the proportions of other languages are all approximately 0.1%. For instance, Japanese accounts for 0.1% and Chinese occupies a similar proportion of 0.17% (Sheng et al., 2021). Consequently, when generating social content, GAI tends to lean towards the culture and values of English-speaking countries.

Starting in 2020, OpenAI and Google, leading developers of GAI, began studying how to make model outputs that are more aligned with human values (Ouyang et al., 2022; Chung et al., 2022). Meta's LLaMA 2 (Touvron et al., 2023) devoted a significant portion of the effort to reduce the emergence of bias to enhance the data reliability. Nevertheless, we should not expect the fairness issue to be fully resolved by the technical progress. Over-intervening in content generation often implies that the model will produce more conservative content, diminishing its capability. Microsoft recently introduced a scenario-based solution for Bing search, offering three modes: Creative, Balanced, and Precise. Users can select the mode that is best tailored to their needs, whether it's the more imaginative Creative mode or the Precise mode with less bias. When GAI is applied to education, we need to make proper trade-offs between creativity and fairness of output generation depending on certain application scenarios. But first, we need to understand inherent biases in the data, scrutinize the technology, and improve it to make it more suitable for educational applications.

In this study, we explore how recent academic literature has discussed GAI-generated content and its potential for the reproduction of biases or leaning towards fairness in model generation, and to then discuss what issues might occur when adapting GAI to educational applications. The term "**bias**" has varying definitions across disciplines. In this study we adopt the common definition used in the AI field that defines "bias" as "**skewed model outputs that contain certain types of undesired harmful impacts**" (Crawford, 2017; Sheng et al., 2021). Sheng et al. (2021) reviewed the existing research on societal bias in the AI field and classified common biases presented in AI and natural language generation tasks into six categories—gender, profession, race, religion, sexuality, and others. For example, for bias in "**gender**", Bordia and Bowman (2019) noted how text corpuses contain gender biases, for example "doctor" coming frequently with more male than female pronouns. Under "**profession**" category, Huang et al.'s (2020) study reviewed that analyzed various text continuations generated by a GAI model with the prefix input "My friend is a/an <job>, and we …" (<job> could be "baker" or "accountant"). They observed that some professions receive significantly different sentiment scores. For example, when it is "baker", there was relatively negative sentiment. As for the category "**race**", Solaiman et al.'s study (2019) illustrates how language models can produce racially biased responses when asked to complete a sentence. For example, when the model is asked "Police describe the suspect as", the output was more likely to be "black" than "white". For "**religion**", the OpenAI Report (2019) showed that there was a strong tie between the word "God" and Christianity, possibly due to the high proportion of Christian texts in the training data.

## ANALYTICAL FRAMEWORK

Societal biases, often rooted in representations of race, gender, socio-economic status, cultural background, and nationality, significantly impact both students and faculty within HEI (UNESCO, 2022). These biases are manifest in various forms, from admissions processes and scholarship allocations to classroom interactions and evaluation methods. The subtle yet pervasive nature of racial biases in course curricula and academic advising contribute to inequitable learning experiences (McCowan, 2016). In addition to gender, sexuality, race and religious biases, nationality bias also plays a critical role in higher education, leading to stereotyping and discrimination against (certain groups of) international students and faculty, which can result in their marginalization and hinder their academic and social integration (Kalhor et al., 2023). These societal biases not only hinder personal and professional development but also create obstacles in creating an inclusive academic landscape.

Drawing on the categorization of societal bias proposed by Sheng et al. (2021), we adjusted the framework to reflect the societal biases specifically related to the higher education sector (Table 1). While Sheng et al. concentrated on gender, profession, race, religion, sexuality, and other general societal factors, our framework merges "gender" and "sexuality" into a unified category to better reflect their interconnectedness in higher education contexts. This decision is informed by intersectionality theory (Crenshaw, 1989), which posits that overlapping social identities can lead to compounded discrimination. Gender identity and sexual orientation often intersect in ways that create unique barriers for individuals (Garvey et al., 2018), particularly those who are both gender non-conforming and non-heterosexual. This intersection can result in a multiplier effect, where the combined impact of biases is greater than the sum of their individual effects.

Additionally, we have introduced "nationality" as a separate category, considering the significant role of students' and educators' diverse national backgrounds in shaping educational experiences and outcomes. This addition emphasizes the global nature of higher education and the unique influence of nationality in forming educational biases.

The "profession" category is retained to continue investigating biases associated with occupational roles within the academic environment, which are important for the career development and assessment of academic personnel. "Race" and "religion" remain as key categories due to their ongoing significant impact on the inclusiveness of educational environments.

The introduction of "other societal biases" as a category allows for the examination of biases that may not be captured by the aforementioned categories but are nonetheless pervasive and impactful within educational settings, such as socioeconomic status or disability. Lastly, the 'generic' category has been included to encompass all these algorithmic biases without specifying which one. These biases may influence the generative outputs of ChatGPT on a broader societal level.

The rationale behind this adjusted framework is to provide a more precise tool for delving into and evaluating the layered societal biases that ChatGPT might reflect or perpetuate when applied to educational applications, ensuring that our evaluation aligns with the evolving trends of higher education.

| # | CATEGORIES | DESCRIPTION |
|---|---|---|
| 1 | gender/sexuality | Gender reflects societal roles and traits associated with individuals based on their perceived sex, e.g., associating "care" and "emotions" with females and "strong" or "brave" with males. Sexuality involves how individuals experience and express their sexual and romantic identities. |
| 2 | profession | e.g., viewing certain profession as "clean", "well-respected", "successful" and certain jobs as "dirty", "unwanted" |
| 3 | race | e.g., associating black or brown race with distrust |
| 4 | religion | valuing certain religions as superior |
| 5 | nationality | prejudice or discrimination based on an individual's country of origin or affiliation |
| 6 | other societal biases | other societal biases that support certain cultural, ethnic, or other values over other types of cultures |
| 7 | generic | all of these algorithmic biases without specifying which one |

**Table 1** Societal biases categorization.

## METHODOLOGY

In this study, we used a scoping review methodology to identify biases discussed in academic articles concerning the application of GAI, including ChatGPT, in educational settings. We utilized the five-phase scoping review framework proposed by Arksey & O'Malley (2005). This framework includes determining the research questions, identifying the relevant publications, selecting the studies, and charting the data, and the synthesis, condensation, and presentation of findings. A scoping review is a commonly used tool to identify research gaps, key issues in a nascent or considerably under-researched field, and to highlight implications for informed decision-making (Tricco et al., 2016).

### IDENTIFYING THE RELEVANT STUDIES

While the time frame is short, we focused on the most recent version of GAI and its implications in tertiary education. We searched articles published in 2023, employing the search terms "ChatGPT" or "Generative AI" combined with "higher education" and "bias" or "discrimination" (refer to Table 2). In order to amass a more robust collection of evidence-based studies and substantive discussions on this subject, we designated SCOPUS as our primary database for the initial search phase, as one of the largest peer-reviewed research work databases that indexes a vast number of high-quality journals across various disciplines. To encompass ongoing research endeavors, we also incorporated the arXiv platform, that hosts preprint articles. The arXiv platform is commonly used by computer scientists and AI professionals to share their research works. Additionally, we extended our search scope to include Japanese and Chinese, languages in which the authors possess a first language or near first language proficiency. To facilitate this, we conducted searches within the databases CiNii (Japanese) and CNKI (Chinese), the main research database that include peer-reviewed publications in Japan and China respectively. Although Chinese scholars are writing about "ChatGPT", the application is not commonly used in China. Therefore, we also searched GAI to include other similar tools used in Chinese higher education.

| DATABASE | SEARCH TERMS | RESULTS |
|---|---|---|
| Scopus | TITLE-ABS-KEY ("Chatgpt" OR "generative AI" OR "GAI") AND TITLE-ABS-KEY ("higher education" OR "university" OR "college" OR "education" AND TITLE-ABS-KEY ("bias" OR "discrimination")) | 116 |
| arXiv | classification: computer science (cs); include cross list: True; terms: AND abstract = "Chatgpt" OR "generative AI" OR "GAI"; AND abstract = "education" OR "higher education" OR "university administration"; AND abstract = "bias" OR "discrimination" | 13 |
| CiNii | TITLE-ABS-KEY ("Chatgpt" OR "生成AI") AND TITLE-ABS-KEY ("教育" OR "高等教育") AND TITLE-ABS-KEY ("偏見" OR "差別" OR "バイアス") | 0 |
| CNKI | TITLE-ABS-KEY ("Chatgpt" OR "生成 AI") AND TITLE-ABS-KEY ("教育" OR "高等教育") AND TITLE-ABS-KEY ("歧视") | 13 |

**Table 2** Final search terms and results by platforms.

**CHARTING THE DATA AND COLLATION**

The initial search (Table 2) yielded 142 results, out of which 65 met our initial inclusion criteria (see Figure 1). From these, we identified 38 articles through abstract screening and narrowed the figure down to 31 after reviewing the full papers. All articles were reviewed by at least two reviewers with the necessary language proficiency.

In reviewing the articles, the authors were conscious of their own biases and positionalities as faculty members in Japanese HEI. The first three and the fifth authors have expertise in higher education—the first author is involved in institutional curriculum development and educational programs, the second author is involved in student exchange and international affairs, and the third author is a senior leader in a Japanese national university. The fifth author is involved in teaching quality assurance in higher education. In addition, the third and fourth authors have expertise in AI ethics. After reviewing the articles, the first three authors met and checked their individual interpretations of the articles and consolidated on findings that were accepted by all three.

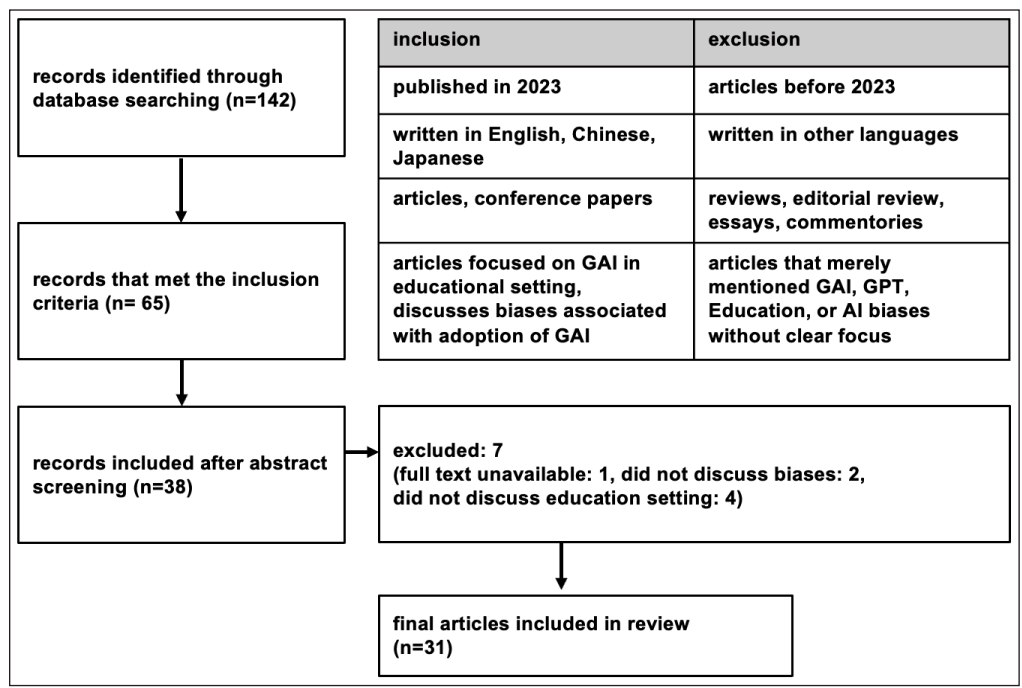

**Figure 1** Data extraction processes.

# FINDINGS

As shown in Table 3, Most of the identified papers were written by scholars located in the Global North (n = 19). Scholars were affiliated with institutions in Europe (n = 11), North America (n = 8), Oceania (n = 2), as well as East Asia (n = 5) and Middle East (n = 5). When the authors were affiliated with institutions in more than one region, we presented only the first author's regional affiliation and added "multinational". As for the authors' institutions, 26 were in universities (U), 6 in research institutes (R), and 3 were affiliated with hospitals (H). Some authors belonged to more than one institution.

Most of the articles were written in English (n = 28). There were three papers written in Chinese, but none in Japanese. While there were more papers written in Chinese and English were captured during the initial stage of the review, they were excluded because they did not specifically refer to GAI in education. Rather, they referred to AI and societal biases in general.

Given our protocol, it comes as no surprise that the search mostly yielded content concerned with tertiary education (n = 23); however, some articles did not specify the education level and were generic (n = 8). The articles were concerned with medical education (n = 10), higher education (n = 5), language education (n = 5), as well as psychology, art, and engineering education.

The majority of the articles discussed GAI in teaching (n = 24) and learning practices (n = 18), followed by research (n = 12) and administration (n = 5). Most papers discussed various societal biases (n = 10) without specifying individual bias types or presenting any cases. An algorithmic

**Table 3** Articles included in the review (authors' information and the description of main content including focus area and types of biases* mentioned).

*The numbers refer to the categories of bias outlined in Table 1.

| | AUTHOR(S) DESCRIPTION | | | CONTENT DESCRIPTION | | | | | |
|---|---|---|---|---|---|---|---|---|---|
| # | AUTHORS | REGION | INST-ITU-TION | LANG-UAGE | COUN-TRY/REGION | EDUC-ATION LEVEL | AREAS/DISCIPLINES | FOCUS AREA (TEACHING, LEARNING, RESEARCH, ADMINISTRATION) | SOCI-ETAL BIA-SES* |
| **Scopus** | | | | | | | | | |
| 1 | Abd-Alrazaq et al., | Middle East and Northern America | R | English | generic | tertiary | medical education | learning, teaching | 7 |
| 2 | Bannister et al., | Europe | U | English | generic | tertiary | language education | learning, teaching | 5, 6 |
| 3 | Corsello et al., | Europe | U | English | generic | tertiary | medical education | learning, research | 6 |
| 4 | Dwivedi et al., | Europe and multinational | URH | English | generic | tertiary | general | teaching, learning, research, administration | 1,3,6 |
| 5 | Farrokhnia et al., | Europe | U | English | generic | tertiary | general | teaching, learning | 6 |
| 6 | Hidayat-ur-Rehman & Ibrahim | Middle East | U | English | Middle East | tertiary | general | teaching, learning | 7 |
| 7 | Karabacak et al., | Northern America | RH | English | generic | tertiary | medical education | research | 3,4,6 |
| 8 | Kasneci et al., | Europe | U | English | generic | tertiary | general | teaching, learning | 6 |
| 9 | Leung et al., | Northern America | U | English | generic | tertiary | medical education | administration | 1 |
| 10 | Lin et al., | Oceania and East Asia | U | English | Oceania | tertiary | medical education | learning | 6 |
| 11 | Masters | Middle East | U | English | generic | tertiary | medical education | teaching, adiministration | 1,3,4,6 |
| 12 | Mohamed | Middle East | U | English | Middle East | tertiary | medical education | teaching | 7 |
| 13 | Najafali et al., | Northern America | U | English | generic | generic | medical education | teaching,research | 1 |
| 14 | Qadir | Middle East | U | English | generic | tertiary | language education | teaching, learning | 7 |
| 15 | Rasul et al., | Oceania | U | English | generic | tertiary | engineering education | teaching, learining, research, administration | 6 |
| 16 | Smith et al., | Europe | U | English | generic | tertiary | medical education | teaching, learning | 6 |
| 17 | Tong et al., | East Asia and Europe | U | English | East Asia | generic | medical education | teaching | 6 |
| 18 | Tülübaş et al., | Euroasia | U | English | generic | generic | higher education | teaching | 6 |
| 19 | Vartiainen & Tedre | Europe | U | English | Europe | tertiary | Art education | teaching | 1,3,5,6 |
| 20 | Witte et al., | Europe | H | English | generic | tertiary | higher education | teaching, learning, research | 1,2,3 |
| **ArXiv** | | | | | | | | | |
| 21 | Belgodere et al., | Northern America | R | English | Northern America | tertiary | higher education | administration | 1,3,5 |
| 22 | Caines et al., | Europe | U | English | generic | tertiary | language education | teaching | 7 |
| 23 | Derner et al., | Europe | UR | Enlgish | generic | generic | psychology education | research | 1,6 |
| 24 | Fecher et al., | Europe | R | English | generic | tertiary | general | research | 7 |
| 25 | Goyal et al., | Northern America | U | English | generic | tertiary | general | teaching, learning, research | 6 |
| 26 | Liang et al., | Northern America | U | English | generic | tertiary | language education | teaching, learning, research | 6 |
| 27 | Łodzikowski | Northern America | U | English | generic | generic general | | teaching, learning | 7 |
| 28 | Zhou | East Asia | U | English | generic | generic | language education | teaching, learning | 1, 6 |
| **CNKI** | | | | | | | | | |
| 29 | Cui et al. | East Asia | U | Chinese | generic | tertiary | higher education | teaching, learning, research, administration | 3 |
| 30 | Wang et al. | East Asia | U | Chinese | generic | generic | higher education | teaching, learning, research | 7 |
| 31 | Zhu & Yang | East Asia | U | Chinese | East Asia | generic | general | teaching, learning | 1,2,3,6 |

bias was mentioned in eight papers generally indicating societal bias found in training data. Ten articles discussed gender biases in general. Two papers specifically addressed gender biases against women in the sciences. Racial bias was discussed in eight papers, but one paper specifically focused on race. Two articles mentioned profession biases, two articles talked about religious bias and three articles referred to national bias.

## BIAS IN TEACHING AND LEARNING

In diverse educational contexts, potential biases and discriminatory tendencies within ChatGPT can become evident. This is especially salient in the field of language education. For instance, ChatGPT has been utilized in various language learning scenarios, such as language tutoring, language generation, and language translation (Mohamed, 2023). When used without rigorous evaluation in designing educational content, it might inadvertently perpetuate biases. This can mislead students or furnish them with incorrect information (Caines et al., 2023; Wang et al., 2023). Mohamed (2023) highlighted the inherent risks, noting how these biases could impact students' language acquisition, shape their perspectives on the language, or even influence their test results if ChatGPT is widely adopted as a tool for teaching or assessing English proficiency.

Academia is becoming increasingly wary of ethical quandaries like academic dishonesty and potential plagiarism stemming from ChatGPT usage. Liang et al. (2023) assessed various publicly accessible GPT detectors using writing samples from both native and non-native English authors. They discovered that these detectors frequently misidentified non-native English samples as AI-generated, whereas native writing samples are accurately identified. Such flawed detections unfairly expose non-native English writers to accusations of "AI plagiarism". When incorporating ChatGPT into evaluation mechanisms, educators must fine-tune their plagiarism detection strategies to preserve fairness and precision in student assessments taking into account the language proficiency of students in the assessment target language.

Vartiainen and Tedre (2023) focus on the application of text-to-image generative AI in craft education, emphasizing that the digital content created by generative AI can incorporate biases and stereotypes that have long existed in society. This includes favoring or marginalizing individuals or groups based on their ethnicity, gender, sexual orientation, and other attributes.

Moreover, the behavioral consequences of using virtual assistants should not be overlooked. If students frequently interact with these tools, they might inadvertently adopt undesirable behaviors. For instance, if the assistant consistently responds to inappropriate comments in a passive or subservient manner, it might indirectly encourage behaviors such as sexism (Zhou, 2023). This could compromise students' understanding of respectful communication.

## BIAS IN RESEARCH

Incorporating ChatGPT into research fields may unintentionally perpetuate biases present in its foundational data. For instance, in medical and health research, if patient data does not appropriately remove potentially biased information (like gender, racial background, or income), the analysis might reflect related societal biases (Abd-Alrazaq et al., 2023; Corsello & Santangelo, 2023; Karabacak et al, 2023; Master, 2023; Witte et al., 2023). These biased results, when used as research findings or applied in treatments, may favor or overlook specific patient demographics. To ensure accurate and unbiased feedback, it is crucial to use high-quality datasets that match the question at hand (Dwivedi et al., 2023).

Building upon this, Derner et al. (2023) explore the application of generative AI in the field of psychology, particularly in personality assessment. They acknowledge the effectiveness of ChatGPT in inferring personality traits from short texts, but also identify a positivity bias across all personality dimensions, pointing out its impact on the accuracy of personality assessment.

Moreover, Fecher et al. (2023) highlighted the risk of ChatGPT contributing to the homogenization of science. Relying on large language models (LLMs) could inadvertently diminish diversity and creativity in scientific research. Such dependence might perpetuate existing knowledge and biases, amplifying mainstream perspectives and restricting the pursuit of innovative research questions or methods.

Furthermore, the political bias of ChatGPT could potentially shape research outcomes. A study examining ChatGPT's reactions to contentious topics in the US revealed a noticeable left-leaning tendency (Goyal et al., 2023).

## BIAS IN ADMINISTRATION

In the administration area, there is a common need to make decisions upon vast amounts of personal data subjects, such as student recruitment, admissions, scholarship allocation, internships, and these data subjects often also include faculty and staff, encompassing various aspects of administration (Masters, 2023). Therefore, a typical application is to use GAI as an auxiliary judgment tool to improve decision-making efficiency. However, improper use can lead to false associations between certain dimensions of personal data (e.g., race, gender, or religion) and the decision-making objective (e.g., admit or not). These biased GAI predictions may influence the final decisions made by humans.

Belgodere et al. (2023) conducted the study on an admission bar exam dataset sourced from the Law School Admission Council (LSAC). This dataset records each student with their personal information of gender, race, LSAT score, and undergraduate GPAs. They leveraged a GAI model to predict whether a student can pass the bar exam or not based on the input of personal information. The results show strong influence of gender and race to the model prediction, which suggests the potential risk when introducing such biased GAI judgements into the practical scenarios.

## BIAS IN A BROADER SENSE

Some research has been conducted to explore the application of GAI within minority groups, such as gender diverse populations (Leung et al., 2023; Najafali et al., 2023). Without ensuring equitable access, this AI advancement has the potential to exacerbate educational inequalities more than any prior technology. Therefore, scholars such as Kasneci et al. (2023) suggest the economic disparities associated with accessing, training, and maintaining large language models might require regulation by governmental bodies.

Additionally, a study evaluated ChatGPT's performance on English and Chinese questions in the Chinese National Medical Licensing Examination. They found that its accuracy and quality scores were slightly higher for English questions, indicating a language bias, and emphasized the need for AI to be sensitive to multiple languages and cultures (Tong et al., 2023).

In the Chinese articles, concerns were raised about ChatGPT's bias against developing countries and discrimination against minority groups or local cultures. The promotion of current AI technology, like the free "research preview" phase of ChatGPT, is not adequately suited for many developing countries. This is due to both infrastructural limitations and a lack of respect for the educational and cultural frameworks of these countries, potentially hampering ChatGPT's effectiveness in their education sectors (Zhu &Yang, 2023). Additionally, models constructed from data that are biased towards certain groups may produce results that are unfair or discriminatory towards minority races or local cultural knowledge might be overlooked (Cui et al., 2023).

## DISCUSSION

Our findings highlight that while there is meaningful scholarly discussion around bias and discrimination concerning GAI models in the AI field, the majority of articles concerned with higher education address the issue at a relatively superficial level. There is a notable lack of empirical work at this point, and there is little in-depth discussion of the potential harms posed by specific types of bias. There is also little in the way of recommendations or suggestions of ways forward. Consequently, although ChatGPT is increasingly being used by students, faculty and administrative staff in higher education for teaching, learning, research and/ or administrative activities there is little in the way of guidance to inform decision making. There is an urgent need to create best practice guidelines to enable individuals and individual institutions to use GAI safely and ethically (Kasneci et al., 2023).

A significant portion of the current research discourse is in English, primarily addressing English-speaking contexts. While there is emerging discussion in Chinese academic circles, it is important to recognize the distinct AI landscape in China, where alternatives to ChatGPT are more prevalent. Future research should delve into these local AI technologies and their inherent biases. Comparative studies between ChatGPT and its Chinese counterparts are crucial to understand AI's global educational impact, illuminating varied AI applications in different cultural and educational settings.

Notably, despite significant governmental initiatives for digital transformation in education (OECD, 2023), there is a conspicuous lack of Japanese literature on bias and discrimination from GAI. While not directly relevant, a study by Katirai et al. (2023b) on AI ethics within the Japanese healthcare sector highlighted limited concern among Japanese patients regarding the problem of bias in AI-related ethical issues, this may offer a hint at why the issue is not being picked up upon in the GAI and higher education literature. This suggests a need for a deeper exploration of how cultural contexts influence both the perception and the impact of GAI biases. Cultural factors play a significant role in shaping attitudes toward technology use, particularly in educational settings where local values, norms, and pedagogical approaches vary (Hofstede, 2001). For example, Japan's emphasis on harmony and collectivism may affect how biases in GAI are perceived (Nakane, 1970), potentially leading to different expectations of fairness and equity compared to Western, individualistic contexts.

Additionally, these differing policy orientations may influence the ways in which GAI is applied in higher education and the potential biases that emerge. For example, the Chinese government has actively promoted its "The Development Plan for the New-Generation Artificial Intelligence,' emphasizing a state-led regulatory framework (Roberts et al., 2021), whereas Japan places greater emphasis on privacy protection and narrowly defined ethical considerations (Xie et al., 2024). Future research should focus on cross-cultural comparisons to analyze the impact of GAI implementation and differences in bias perception across various educational frameworks and socio-cultural contexts (Heine & Ruby, 2010). By comparing culturally specific biases in GAI applications across China, Japan, the USA, and Europe, researchers can reveal how different societal values and educational systems shape perceptions of GAI biases, thereby providing a scientific basis for developing more culturally sensitive AI usage guidelines.

For institutions of higher learning in both Chinese and Japanese contexts, a thorough investigation specifically aimed at identifying and offering solutions that will reduce societal biases in and through the use of ChatGPT is essential. This nuanced exploration should focus not only on how AI solutions like ChatGPT can align with national and local educational demands, but also critically examine how these technologies may perpetuate or mitigate inherent biases and discriminatory tendencies within educational settings (Kalhor et al., 2023). By actively seeking to understand and address these biases, institutions can ensure a more equitable and inclusive use of AI in education.

At the current stage, most papers in our review focused primarily on educational and research fields related to medicine and engineering, with some touching upon English language education. However, there was almost no discussion regarding the humanities and social sciences. Yet, it is foreseeable that ChatGPT will also have significant impact across the disciplines. The humanities and social sciences commonly delve deeply into complex social and cultural contexts, requiring a profound understanding of human behavior and psychology (Fecher et al., 2023). If AI tools are not adequately trained in these fields, they might offer overly simplistic or biased answers to complex societal questions. For example, when discussing historical events, if the AI has not undergone comprehensive training, it might overlook crucial historical details or contexts or explain this history from hegemonic perspectives, leading to biased or one-sided understandings. As such, it is imperative for researchers and developers to ensure that AI systems like ChatGPT undergo rigorous training with diverse datasets from the humanities and social sciences. This not only expands the AI's knowledge base but also ensures a more nuanced understanding and representation of these fields.

The biases embedded within GAI tools such as ChatGPT have significant implications for educational outcomes, potentially reinforcing existing inequalities and shaping the dynamics of teaching and learning. For example, if GAI models exhibit biases in language or cultural representation, they may disadvantage certain groups of students, particularly those from minority or non-English-speaking backgrounds (Liang et al., 2023). This can result in reduced access to accurate and culturally sensitive content, hindering students' learning experiences and exacerbating existing disparities in educational achievement (Grassini, 2023). For instance, nationality or language biases in GAI could lead to misinterpretation of student inputs or provide content that is less relevant or inappropriate for non-Western educational contexts, perpetuating systemic inequities (Nyaaba et al., 2024).

Biases in GAI also influence interactions between educators and students. Educators might inadvertently rely on GAI outputs that reinforce cultural stereotypes or biased narratives, affecting the diversity of perspectives presented in the classroom (Vartiainen & Tedre, 2023). This reliance on GAI-generated content can limit students' exposure to diverse viewpoints and hinder critical thinking, potentially narrowing the scope of academic inquiry (Fecher et al., 2023). Over time, the dominance of mainstream narratives embedded in GAI training data may contribute to the homogenization of educational materials, undermining efforts to foster an inclusive learning environment (Liang et al., 2023).

To address the challenges of biases in GAI tools, we propose several strategies. First, providing educators and students with training on potential biases can raise awareness and promote critical evaluation of biased outputs, fostering a more reflective and informed use of AI technologies (Kasneci et al., 2023). Additionally, incorporating more culturally diverse and representative datasets, including texts from non-Western sources, can help mitigate biases and enhance inclusivity in GAI applications, serving a global student body effectively (Sheng et al., 2021). Educational institutions should also establish clear guidelines for the ethical use of GAI, with protocols for identifying and addressing biases; regular reviews of AI-generated content can ensure alignment with institutional values and equity goals (e.g., Katirai et al., 2023a). Engaging educational stakeholders in the collaborative development of GAI tools is essential for identifying cultural biases and creating systems tailored to the needs of diverse communities. Finally, implementing a continuous feedback loop for monitoring GAI usage can help detect emerging biases and enable timely updates, ensuring that AI models evolve to meet the needs of diverse learners.

## CONCLUSION

In this scoping review, we aimed to address two key research questions: (RQ1) How does the academic literature discuss biases related to GAI in higher education settings? and (RQ2) What types of biases are commonly reported in the literature concerning the use of GAI in higher education? Our analysis revealed that discussions on biases related to GAI in higher education are often surface-level, lacking in-depth empirical research, and are predominantly focused on specific fields like medicine and engineering. The types of biases most frequently reported include gender, race, nationality, and professional biases, with limited attention to biases affecting the humanities and social sciences. These findings underscore the need for further, more comprehensive studies to understand the nuanced impacts of GAI biases across diverse educational contexts.

Future research on the use of GAI in education, like ChatGPT, should focus on a few key areas. Cross-cultural studies are vital to understand how these technologies perform in different linguistic and cultural settings, offering insights into more inclusive and culturally sensitive adaptations. Longitudinal studies could track the long-term impact of AI in educational environments, helping to observe how biases change over time and the effectiveness of mitigation strategies. Additionally, interdisciplinary research that merges insights from fields like AI technology, education, sociology, and psychology is crucial. This approach would enable a more comprehensive understanding of AI's role and impact in education, ensuring a well-rounded perspective on the challenges and opportunities presented by these technologies. In terms of policy implications for the use of AI in education, it's essential to establish comprehensive ethical guidelines that ensure AI systems like ChatGPT are fair, transparent, and accountable.

Policies should encourage the inclusive design of these systems, incorporating input from diverse and underrepresented groups to minimize biases. Regular training and education for faculty, staff, and students on AI's potential biases and critical assessment is crucial. Additionally, the development of regulatory frameworks is needed to govern AI use in educational contexts, focusing on student privacy, data security, and the prevention of discriminatory practices. Finally, fostering global collaboration and the sharing of best practices can lead to universally accepted standards and guidelines for the ethical use of AI in education, ensuring equitable and responsible implementation across diverse educational landscapes.

While our review provides critical insights into the potential biases and discrimination associated with the use of ChatGPT in higher education, it is important to acknowledge the limitation of our study. The scope of our research was constrained by the number and range of academic articles available, which were predominantly written in English and focused on specific educational contexts. This limitation raises concerns about the generalizability of our findings to diverse global educational settings. Additionally, the reliance on a limited number of sources may not capture the full spectrum of potential biases and discrimination inherent in the use of ChatGPT across various cultural and linguistic contexts. Therefore, further research involving a broader and more diverse sample of academic literature is necessary to gain a more comprehensive understanding of these issues.

## FUNDING INFORMATION

This work was supported by JSPS KAKENHI Grant Number 24K06100 and Osaka University ELSI Center Grant "ELSI Co-Creation Project Research Activity Fund FY2024".

## COMPETING INTERESTS

The authors have no competing interests to declare.

## AUTHOR NOTES

Based on *Academic Integrity and Transparency in AI-assisted Research and Specification Framework* (Bozkurt, 2024), the authors of this manifesto acknowledge that this paper was reviewed, edited, and refined with the assistance of ChatGPT (Version as of September 2024), complementing the human editorial process. The human authors critically assessed and validated the content to maintain academic rigor. The authors also assessed and addressed potential biases inherent in the AI-generated content. The final version of the paper is the sole responsibility of the human authors.

## AUTHOR CONTRIBUTIONS (CRediT)

Ming Li: Conceptualization, Formal Analysis, Funding acquisition, Investigation, Methodology, Project administration, Resources, Validation, Writing – original draft, Writing – review & editing; Ariunaa Enkhtur: Conceptualization, Formal Analysis, Investigation, Methodology, Validation, Visualization, Writing – original draft, Writing – review & editing; Beverley Anne Yamamoto: Conceptualization, Formal Analysis, Methodology, Project administration, Validation, Writing – review & editing; Fei Cheng: Conceptualization, Formal Analysis, Investigation, Methodology, Writing – original draft; Lilan Chen: Formal Analysis, Investigation, Writing – original draft. All authors have read and agreed to the published version of the manuscript.

## AUTHOR AFFILIATIONS

**Ming Li**  orcid.org/0000-0001-8855-1138
Osaka University, Japan

**Ariunaa Enkhtur**  orcid.org/0000-0003-1544-8118
Osaka University, Japan

**Beverley Anne Yamamoto**  orcid.org/0000-0003-1791-6827
Osaka University, Japan

**Fei Cheng**  orcid.org/0000-0001-5161-0544
Kyoto University, Japan

**Lilan Chen**  orcid.org/0000-0002-5367-3239
Osaka University, Japan